\documentclass[12pt]{iopart}

\newcommand{\gF}{\mbox{$g_{\rm{{F}}}$}}
\newcommand{\mF}{\mbox{$m_{\rm{{F}}}$}}
\newcommand{\muB}{\mbox{$\mu_{\rm{{B}}}$}}
\newcommand{\omegaR}{\mbox{$\omega_{\mbox{ \textrho}}$}}
\newcommand{\omegaZ}{\mbox{$\omega_{\rm{z}}$}}
\newcommand{\omegaRF}{\mbox{$\omega_{\rm{rf}}$}} 
\newcommand{\OmegaRF}{\mbox{$\Omega_{\rm{rf}}$}}
\newcommand{\OmegaL}{\mbox{$\Omega_{\rm{L}}$}}
\newcommand{\unit}[1]{\ensuremath{\, \mathrm{#1}}}

\usepackage{
     hyperref, 
     textcomp, 
     textgreek, 
     todonotes 
     }

\begin{document}


\title{An ultra-bright atom laser}

\author{V.\,Bolpasi$^{1,2}$,
N.K.\,Efremidis$^3$, 
M.\,J.\,Morrissey$^1$\footnote[1]{Quantum Research Group, University of KwaZulu-Natal, Westville, Durban 4000, South Africa}, 
P.\,Condylis$^1$\footnote[2]{Centre for Quantum Technologies, National University of Singapore, 3 Science Drive 2, Singapore 117543, Singapore}, 
D.\,Sahagun$^1$\footnotemark[2], 
M.\,Baker$^1$\footnote[3]{School of Mathematics and Physics, The University of Queensland, Brisbane QLD 4072, Australia} 
and W.\,von\,Klitzing$^1$
}

\address{$^1$ Institute of Electronic Structure and Laser, Foundation for Research and Technology - Hellas, P.O. Box 1527, 71110 Heraklion, Greece}

\address{$^2$ Physics Department, University of Crete, 71103 Heraklion, Crete, Greece}

\address{$^3$ Applied Mathematics Department, University of Crete, P.O.\,Box 2208, 71003 Heraklion, Greece}

\ead{{\mailto{AtomLaser@bec.gr}} /  \url{http://www.bec.gr}}

\date{\today}

\begin{abstract}
We present a novel, ultra-bright atom-laser and ultra-cold thermal atom beam. 
Using rf-radiation we strongly couple the magnetic hyperfine levels of $^{87}$Rb atoms in a trapped Bose-Einstein condensate. 
The resulting time-dependent adiabatic potentials forms a  trap, which
at low rf-frequencies opens up just below the condensate and thus allows an  extremely bright well-collimated  atom laser to emerge. 
As opposed to traditional atom lasers based on weak coupling of the magnetic hyperfine levels, this technique allows us to outcouple atoms at an arbitrarily large rate. 
 We achieve a flux of $4\times10^7$ atoms per second, a seven fold increase  compared to the brightest atom lasers to date. 
 Furthermore, we demonstrate by two orders of magnitude the coldest thermal atom beam (200\,nK).
\end{abstract}


\maketitle

\section{Introduction \label{sec:Intro}} 

\setcounter{footnote}{0}
Atom laser beams are coherent matter waves outcoupled from of a Bose-Einstein condensate (BEC).  
 They are the matter-wave analogue to the photon laser, with the magnetic trap corresponding to the optical cavity and the freely propagating atoms to the laser beam. 
Soon after the demonstration of BEC in dilute gases \cite{Davis1995PRL,Anderson1995S}, researchers demonstrated the first output couplers for matter-waves \cite{Mewes1997PRL, Anderson1998S, Hagley1999S, Bloch1999PRL,Cennini2003PRL}.
 A series of beautiful experiments measured the spatial phase coherence \cite{Bloch2000N} and the atom number correlations \cite{Ottl2005PRL,Kohl2007APB}. 
Matter waves have been coupled into the fundamental mode of matter-wave guides \cite{Guerin2006PRL,Dall2010PRA}, and
beam splitters \cite{Gattobigio2012PRL} and Bragg reflectors \cite{Fabre2011PRL} have been  demonstrated.
Recently, the arrival of a purely laser-cooled BEC has opened the possibility of a truly continuous atom-laser \cite{Stellmer2013PRL}.
In addition to their fundamental interest there are promising applications for bright coherent atom lasers.
They would allow higher sensitivity in matter-wave interferometry \cite{Dowling1998PRA,Berman1996BOOK,Cronin2009RMP}, which could be further enhanced by number squeezed atom laser beams \cite{Haine2005PRA,Johnsson2007PRL,Cronin2009RMP}. 
The extremely tight focal spot of a coherent atom lasers could be exploited in direct atom lithography \cite{Shvarchuck2002PRL} or ultra-sensitive magnetometry with very high spatial resolution \cite{Vengalattore2007PRL}. 
An extremely cold thermal atom beam would be very useful for high-resolution spectroscopy of ultra-cold collisions. A detailed review of the state-of-the-art can be found in \cite{Robins2013PRSOPL}.

A key factor in the performance of all of these devices is the spectral brightness and therefore the flux of the atom laser beam. In this respect the atom laser is superior to the thermal atom beam like the optical laser to the ordinary light bulb. 
The flux of the traditional atom-laser achievable from a given condensate has been fundamentally limited by the outcoupling process. 
In this paper we present a novel output-coupler for magnetically trapped atoms, which eliminates this limit. We demonstrate an increase in the flux of the atom laser by a factor of seven compared to the brightest atom lasers demonstrated so far. This novel output coupler allowed us also to produce a thermal beam, which is by two orders of magnitude the coldest every observed.

\begin{figure}\center
 \includegraphics[width=0.5 \columnwidth]{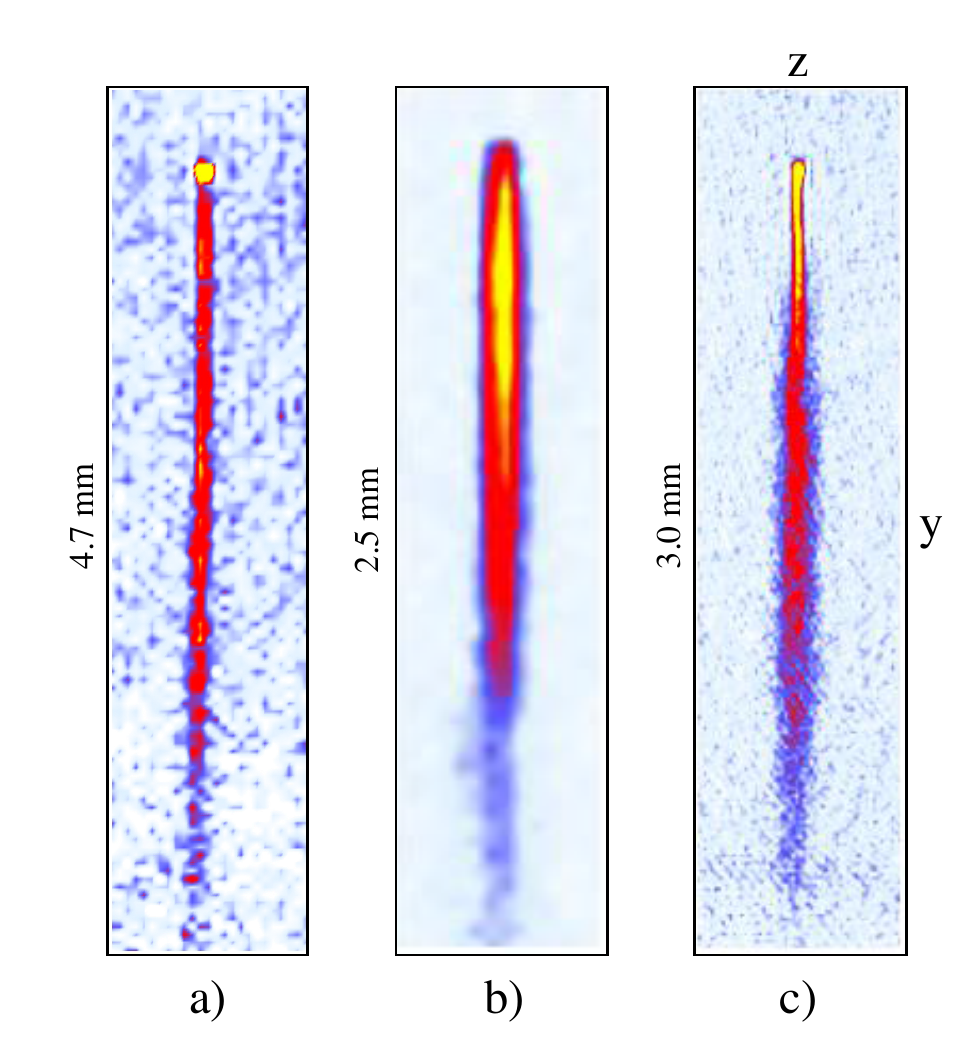}%
 \caption{Atom density distributions of atom-lasers:
 a)\ A well collimated atom laser beam, 
b)\ an ultra-high flux atom laser, 
and c)\ atom beam combining atom laser and thermal emission. 
The density distributions were measured by absorption imaging after time-of-flight expansion. The experimental parameters for all three experiments were $\alpha=440\,{\rm G/cm},\beta=170\,\rm{G/cm}^2$. The radio frequency was ramped at a speed of $\dot\omega_{\rm{rf}}/2\pi=5$\,MHz/s. The offset field $B_0$ was 1\,G, 0.5\,G and 0.5\,G respectively. The duration of the output coupling of the atom beams was 2\,ms, 1\,ms and 10\,ms. (Please note the different vertical scales in the three plots.)
 \label{fig:AtomLasers}}
 \end{figure}

\section{Atom Laser Output Couplers \label{sec:OutputCouplers}} 
\subsection{Output coupling using weak fields \label{sec:WeakOutputCoupling}}
Atom laser beams can be generated in a number of different ways: Pulsed atom lasers have been produced from magnetically trapped BECs by short blasts of rf-radiation\,\footnote{The rf in \cite{Mewes1997PRL} was  pulsed in intensity or swept in frequency at a rate of 500\,MHz/s, which means that in both cases the atoms released  on a time scale much shorter than a quarter oscillation of the atoms in the magnetic trap.} \cite{Mewes1997PRL}. 
An atom laser based on spilling atoms from a purely optical potential has been demonstrated for small laser fluxes \cite{Cennini2003PRL}. 
Bright quasi-continuous atom lasers are usually outcoupled from magnetically trapped BECs using weak electromagnetic fields. This couples the Zeeman levels of the atoms and thus the trapped states to propagating ones. 
The transition can either be direct (rf output coupler) \cite{Bloch1999PRL} or via a virtual state (Raman output coupler) \cite{Hagley1999S}.
Even though this process is irreversible \cite{Debs2010PRA}, it is coherent in the sense that the atoms retain a well determined phase relationship with the other atoms in the beam and in the condensate \cite{Kohl2001PRL}. 
After the atoms have been transferred to the propagating states they accelerate out of the condensate and form the atom laser beam. 
This acceleration is due to a combination of gravity, magnetic forces, and interaction with the atoms remaining in the condensate.
For very weak fields, the flux of the atom laser is proportional to the intensity of the rf-field \cite{Gerbier2001PRL}. 
For stronger coupling fields, however, a bound state appears, which eventually shuts off the lasers thus severely limiting the maximal atom flux that can be achieved from a given condensate \cite{Jeffers2000PRA,Robins2005PRA}. 

In the weakly coupled atom laser, the outcoupling occurs wherever the rf frequency is resonant with the local magnetic field.  If this happens somewhere inside the condensate, then the chemical potential of the remaining condensate will act as as a matter-wave lens, which can lead to sever distortions of the atom-laser beam   \cite{Riou2006PRL}.

\subsection{Output coupling using strong fields \label{sec:StrongOutputCoupling}}

Here we present a novel type of continuous atom laser, which is based on time-dependent adiabatic potentials (TDAP). Rather than using a weak rf-field to slowly outcouple a small fraction of the condensate, we use strong rf-fields to deform the trapping potential such that the  condensate spills slowly out of a small opening at the bottom of the trap. 
As opposed to the atom laser based on the weak-field output coupler, the
TDAP can emit atoms from the condensate at an almost arbitrary rate.
This allowed us to generate atom-laser beams, which have a much larger flux than can be achieved in the weak coupling regime. 
The TDAP also permitted us to produce the coldest atom beam observed so to date.
Examples  can be seen in Fig.\,\ref{fig:AtomLasers}: The central panel shows an extremely bright atom laser, the left panel  a highly collimated atom laser, and  the right image for the first time a combined thermal and atom-laser beam. 

\section{ Theory \label{sec:Theory}}

\begin{figure}\center
 \includegraphics[width=0.8 \columnwidth]{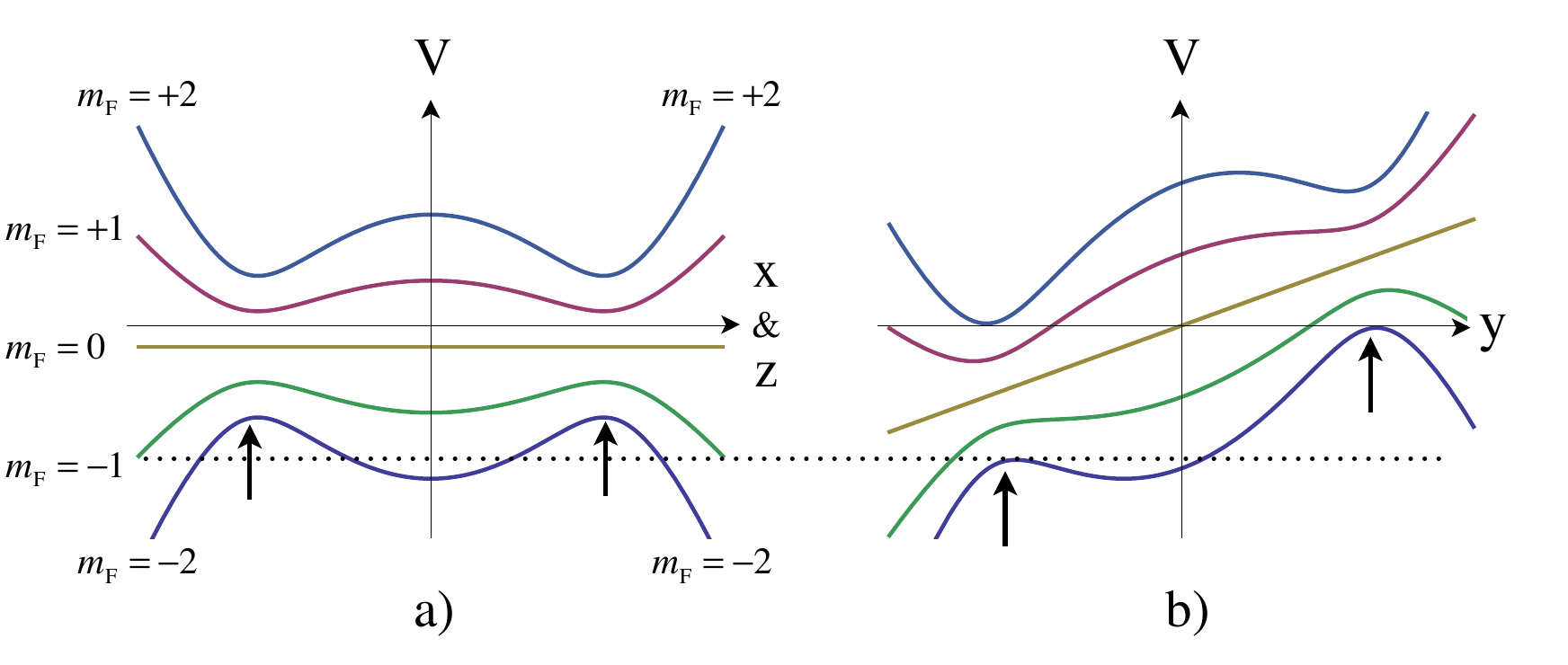}%
 \caption{The dressed trapping potentials in the presence of gravity. 
a) depicts the potential in the horizontal directions (not to scale).
b) shows the potentials in the direction of gravity. 
 The dotted line denotes the energy, above which gravity pulls the atoms out of the trap. 
The arrows indicate the positions, where the radio frequency is resonant with local static B-field, i.e. where $\omegaRF=\OmegaL $.\label{fig:potentials1D}}
 \end{figure}

The adiabatic potentials of the TDAP atom laser are created by subjecting magnetically trapped atoms (here $^{87}$Rb, F=2, m$_{\rm F}$=-2) to a strong rf-field.  Atoms, which pass  through the region where they are resonant with the rf-field, are adiabatically transferred from their trapped to their anti-trapped state, thus limiting the trap depth (Fig.\,\ref{fig:potentials1D}).  At higher rf-frequencies this is used in forced evaporative cooling, where one removes the hottest atoms of a trapped cloud by slowly ramping down the rf-frequency.  At lower rf-frequencies the gravitational tilt of the potential causes the atoms to escape preferentially downwards producing to a thermal atom beam. 
 As the trap depth approaches the chemical potential of the BEC, an atom laser beam emerges from the bottom edge of the condensate.
Since the atoms are adiabatically transferred from the trapped to the anti-trapped state, all atoms of the BEC enter the atom laser beam. The flux of the TDAP atom laser is therefore simply a function of the rate at which the rf-frequency is ramped down\footnote{Note that as opposed to the TDAP, the weakly coupled regime transfers the atoms irreversibly into a number of different magnetic hyperfine states \cite{Debs2010PRA}.}.

We focus our description of TDAP atom lasers  on cigar-shaped Ioffe--Pritchard type magnetic traps. If the radial trapping frequency  $(\omegaR)$ is much larger than the axial one $(\omegaZ)$, then the magnetic field of a Ioffe--Pritchard trap can be written as 
$ {\bf B}\left ({\bf r}\right) =\left\{\alpha\, x, -\alpha\, y, B_0+ \frac{1}{2}\,\beta z^2\right\} 
$,
 where $\alpha$ is the gradient of the radial quadrupole field and $\beta$ is the curvature of the axial field. The 
Larmor frequency associated with the difference in energy between adjacent Zeeman levels is $\OmegaL =\left| \gF \muB {\bf B} ({\bf r}) \right|/\hbar$, where $\gF$ is the Land\'e $g$-factor of the considered hyperfine manifold and $\muB$ is the Bohr magneton. 
We couple the magnetic hyperfine states using a linearly polarised oscillating magnetic field of strength ${\bf B}_{\rm{rf}}$ and  angular dressing frequency $\omegaRF$. The resulting rf coupling strength is $\OmegaRF=|\gF \muB/ 2\hbar|\cdot 
\left|\,{\mathbf{B} (\mathbf{r})}/{\left |\mathbf{B} (\mathbf{r}) \right|} \times
\mathbf{B}_{\rm{rf}}\,\right|$. 

In the weakly coupled atom laser the atomic motion quickly renders irreversible the transfer of the atoms to other spin states \cite{Debs2010PRA}.
In the TDAP atom laser, however,  the coupling strength is large compared to the change in Larmor frequency due to the motion of the atoms $(\mbox{$\Omega^2_{\rm{rf}}$} \gg \dot \Omega_{\rm L}/2\pi)$. Therefore, the spin-transfer is fully adiabatic. We can then ignore the external degrees of freedom and use dressed states to describe the atom in the fields \cite{Lesanovsky2007PRLa, Zobay2001PRL}. 
Taking into account gravity we can write the adiabatic potential as \cite{Lesanovsky2007PRLa,Lesanovsky2006PRA2,Zobay2001PRL}:
\begin{equation}\label{eq:potenital}
V ({\bf r})= \mF \hbar \,\sqrt{\left ( \Omega _{\rm L} - {\omega _{\rm {rf}}}\right)^2+ \Omega _{\rm{rf}}^2}+M g_{\rm e} y \,,
\end{equation}
where $M$ is the mass of an atom, $g_{\rm e} $ earth's gravitational acceleration, and $\mF$ the magnetic quantum number of the total atomic spin when $\Omega _{\rm L}$ is large $(\Omega _{\rm L} \gg \omegaRF,\OmegaRF)$.

\begin{figure}\center
 \includegraphics[width=0.73 \columnwidth]{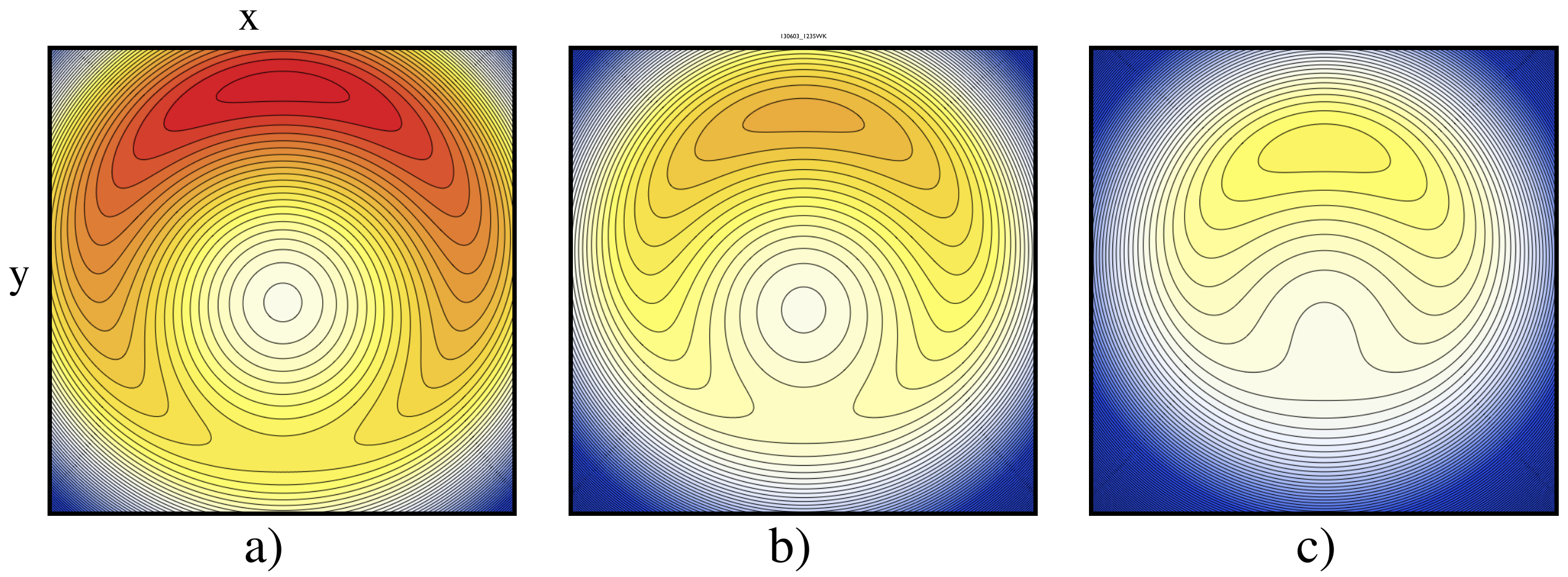}%
 \caption{Contour plot of the dressed trapping potentials in the y\mbox{-}x plane for $m_{\rm F}$=$-$2. The trap parameters are the ones of Fig.\,\ref{fig:AtomLasers}a. The three plots have decreasing rf-frequencies $(\Delta \omegaRF/2\pi)$ with a) being the deepest and in c)just about ceasing to exist. In a) and b), the rf-frequency is 30\,kHz and 15\,kHz above the one of c).  The equipotential lines are spaced by 100\,nK and the colour scale is relative to the trap bottom. Note that the potential in the y\mbox{-}z plane is very similar to the one displayed here, except for a much elongated z-axes.
 \label{fig:potentialsContour}} 
 \end{figure}

Fig.\,\ref{fig:potentials1D} shows the dressed Zeeman states of Eq.\,(\ref{eq:potenital}) for a spin two particle in a harmonic magnetic trap. The left panel shows the potential in the horizontal (x or z) direction and the right panel in the vertical (y) direction. The arrows indicate the position where the rf is resonant with the local magnetic field. 
 The maximum energy of a trapped condensate (the minimum energy, at which the atoms escape into the atom laser beam) is indicated by the dotted horizontal line. 
 Fig.\,\ref{fig:potentialsContour} shows a contour plot of the trapping potentials relative to the trap bottom\footnote{The {\it rf-frequency relative to the trap bottom} is the rf-frequency of the dressing field minus an offset, which is chosen such that at zero trap depth the {\it rf-frequency relative to the trap bottom} is zero.}
 for the $m_{\rm F}=-2$ state for three different values of $\omegaRF$\,:\, 
 a) where the rf-frequency is well above the trap bottom and the trap is relatively deep, 
 b) where the rf-frequency is smaller and the trap more shallow, and 
 c) where the rf-frequency is such that the trap just ceases to exist.

The trap depth can be adjusted dynamically simply by changing the value of $\omegaRF$\,: As one lowers the rf frequency, atoms which had been trapped at an energy larger than the new trap depth escape at the bottom of the trap.
The rate at which one can change $\omegaRF$ --- and thus the rate at which atoms are outcoupled --- is only limited by the requirement that the spin flips be adiabatic. This limits the ramping rate of the rf-frequency to $\dot\omega_{\rm rf}\ll \Omega_{\rm rf}^2$. For our typical coupling strength of $\OmegaRF/2\pi=16$\,kHz this imposes  $\dot\omega_{\rm rf}/2\pi\ll1.6\times10^{9}$\,Hz/s. A condensate with a typical chemical potential of the order of kHz could therefore be released within $1\,$\textmu s, which is much faster than the radial oscillation time. This stands in stark contrast to the weakly coupled atom laser, where the outcoupling rate is limited by the appearance of bound states \cite{Debs2010PRA}.

\subsection{Shape of the atom laser beam \label{sec:Shape}}

The exact shape of the outcoupled atom beam is difficult to predict analytically. 
However, we can distinguish different regimes of output coupling  by comparing the trap frequencies $(\omegaZ, \omegaR)$ to the output coupling rate of the condensate $(\Omega_{\rm oc})$, which represents the inverse of the time that it takes to outcouple the entire condensate.
 At output coupling rates well below the axial trapping frequency $(\Omega_{\rm oc} \ll\omegaZ\ll\omegaR)$, the shape of the condensate can adapt adiabatically to the reducing atom number \cite{Vermersch2011PRA}. The output coupling will then only occur in a very small region below the centre of the condensate. 
At larger outcoupling rates $(\Omega_{\rm oc} \simeq\omegaZ\ll\omegaR)$ the outcoupling region  grows larger and shape oscillations are excited in the condensate, which affects the intensity and the shape of the atom laser beam.
 When the output coupling rate is large compared to the axial frequency but is still small compare to the radial trapping frequency $(\omegaZ\ll\Omega_{\rm oc}\ll\omegaR)$, then the shape in the axial direction remains essentially frozen \cite{Shvarchuck2002PRL,Buggle2005PRA} and  the output coupling will occur below the whole length of the cigar-shaped condensate. 
 Finally, when the sweep of the rf-frequency is fast compared to the radial trapping frequency $(\omegaZ \ll \omegaR \ll \Omega_{\rm oc} )$ then the atoms will form a single accelerating shell-shaped matter-wave pulse \cite{Mewes1997PRL}.

A number of factors lead to the atom laser being very well collimated both in the slow  $(\Omega_{\rm oc} \ll\omegaZ\ll\omegaR)$ and intermediate $(\omegaZ \ll \Omega_{\rm oc}  \ll \omegaR )$ outcoupling regimes: The absence of matter-wave lensing, the smooth shape of the  potential, the absence of shape oscillations, and the vertical acceleration of the atoms after they leave the trap: 
Matter wave lensing occurs in  atom lasers based on a weak rf-field. When the laser beam is outcoupled closer to the centre of the condensate  the chemical potential of the remaining condensate acts as a matter-wave lens, which distorts the atom laser beam.
For the TDAP atom laser, however, matter wave lensing is completely absent  because the atoms are outcoupled from the very edge of the condensate where the chemical potential is negligible.
Furthermore, the shape of the confining potential at the outcoupling point  tends to be very smooth  in the direction transverse to the atom beam and the transverse gradient and curvature of the adiabatic potential become vanishingly small for very low output coupling rates $(\Omega_{\rm oc} \ll\omegaZ \ll \omegaR)$. 
Shape oscillations of the condensates are absent because in slow and intermediate outcoupling regimes the internal dynamics of the condensate are negligible. 
Finally, since the outcoupled atoms are in the anti-trapped magnetic hyperfine state  they are strongly accelerated downward, thus minimising the influence of the transverse momentum component.

 Numerical simulations of the time-dependent Schr\"odinger equation confirm that the TDAP atom laser can produce well-collimated quasi-continuous atom-laser beams (see appendix \ref{sec:numerics}).

\subsection{Longitudinal velocity \label{sec:LongitudinalVelocity}}
For many applications it is desirable for the atom laser beam to have a low transverse and a large longitudinal velocity. Examples include collision physics \cite{Buggle2004PRL, Thomas2004PRL}, surface science\cite{Pasquini2004PRL, Druzhinina2003PRL} and atom nano-lithography \cite{Juffmann2009PRL}. 
The acceleration of  the TDAP atom laser due to the atoms being in the anti-trapped state $(m_{\rm F}=-2)$ provides these high velocities, reaching 2\,m\,s$^{-1}$ after only 1\,cm of travel.

For other applications a slower beam might be more useful.  In this case it is sufficient to introduce a second rf-field, which transfers the atoms from the $m_{\rm F}=-2$ back to the $m_{\rm F}=+2$ state,  which causes the atom laser beam to decelerate and eventually  come to a standstill thus providing access to a very large range of longitudinal velocities \cite{Bloch2001PRL}.

\subsection{Image analysis \label{sec:Fitting}}
Absorption images of the atom laser in the y-z plane are taken by shining resonant laser light along the x-axes, image it onto a CCD camera, and calculate the atom column density for each pixel.
We then cut the images into horizontal slices, integrate each slice in the vertical direction, and fit them with the following binomial profile:
\begin{equation}
f=a + b\,e^{ - \left (\frac{x - x_0}{\Delta x_{\rm t}}\right)^2}\!+ c\,\rm{Re}
{{{\left[ {{{1 - }}{{\left ( {\frac{x - x_0}{{\Delta x_{\rm{c}}}}}\right)}^2}}\right]}^{{{3/2}}}}}.
\label{eq:fitFunction}
\end{equation}
The first term accounts for any global offset. The second term represents the Gaussian momentum distribution of a thermal contribution. The third term has been chosen \emph{ad-hoc} 
for the apparent absence of wings in some of some of the experimental images of the atom laser. It implies an inverted-parabola shape of the transverse density profile , which is integrated in the direction of the imaging beam. Examples of such fits can be found in the supplementary material.

From the fits to the individual integrated atom slices we can then determine the atom number and sizes of both the coherent and thermal components. 
From the position of a slice we can calculate the velocity and output coupling time of the atoms in it and subsequently the divergence, temperature, and local flux of the atom beam. If no reliable fit can be obtained for the bimodal fit, we force either $b=0$ or $c=0$ in order to obtain a fit  for only an atom-laser or only a thermal beam respectively.

\section{Experiment\label{sec:Experiment}} 

\subsection{Making the atom-laser and BEC}
 The creation of the TDAP atom laser is rather straight forward: A  linear sweep of the rf-frequency cools a thermal cloud of magnetically trapped atoms into a BEC and then outcouples the atom laser from the trap. 
 
 Our experimental apparatus has been described elsewhere \cite{Bolpasi2012JOPB, Sahagun2013OC}. We load $10^{9}$ atoms into a shallow magnetic Ioffe Pritchard trap. 
Within one second we compress this trap to its final state by ramping up the currents in the Ioffe and pinch coils whilst simultaneously reducing $B_0$ to 0.5--1\,G. The final trapping frequencies are then 17\,Hz in the axial  and 561--793\,Hz in the radial direction. Then we lower the rf-frequency down to its final value in a linear ramp of 10\,s duration (Fig.\,\ref{fig:potentialsContour}a-c). 
The sample cools by forced evaporation until the critical temperature is reached and a BEC of about $1.5\times 10^5$ atoms forms. 
A thermal atom beam emerges from the bottom of the trap. 
As the rf-frequency drops even lower, the trap depth becomes smaller than the chemical potential of the condensate and an atom-laser is outcoupled from the lower edge of the condensate (Fig\,\ref{fig:potentialsContour}b) until finally the trap vanishes (Fig.\,\ref{fig:potentialsContour}c). 
We then switch off the magnetic trap, let the atoms evolve in free flight for 1-10\,ms, and take a resonant absorption picture of y-z plane. 

Note that the requirements on the reproducibility of $B_0$ are relatively low: Whereas the traditional atom laser needs to aim the weak rf exactly to the bottom of the BEC, the TDAP atom-laser only needs to sweep across it. In our case the magnetic field reproducibility is of the order of a few milli-Gauss with a shot-to-shot noise of less than one milli-Guass and a 50\,Hz modulation of 40\,mG. 
 
\subsection{Atom Lasers \label{sec:experimentalResults}}

In this section we describe in detail  three different atom lasers: A well-collimated pure atom laser (Fig.\,\ref{fig:AtomLasers}a), an atom laser with a very high atom-flux (Fig.\,\ref{fig:AtomLasers}b), and finally an atom beam containing at the same time an atom laser and a thermal atom beam (Fig.\,\ref{fig:AtomLasers}c).
The trap had a gradient $\alpha=440\,{\rm G/cm}$ and a curvature $\beta=170\,\rm{G/cm}^2$. 
The rf-frequency was ramped down from 50\,MHz at a speed of $\dot\omega_{\rm{rf}}/2\pi=5$\,MHz/s. 
The output coupling was therefore  at an intermediate rate  $(\omegaZ\ll\Omega_{\rm oc}<\omegaR)$.
We generate the rf with a direct digital synthesis card (Analog Devices AD9854/PCBZ), amplify it using an rf amplifier (Amplifier Research 25A250A), which is coupled to two antennas in parallel (4\,cm diameter, three windings each).
 
We determine the absolute coupling strength $\OmegaRF$ by measuring the rf-frequencies $\omegaRF_{1,2}$ at which the trap vanishes for two different coupling strengths $\OmegaRF_{1,2}$.
Knowing the \textit{relative} coupling strength $\OmegaRF_2 / \OmegaRF_1 = 10$ and \textit{difference} frequency at which the traps vanish $(\omegaRF_2- \omegaRF_1)/2\pi=16$\,kHz., we can read the coupling strength off Fig.\,\ref{fig:trapDepth2D} (p.\pageref{sec:VelocityMapping})  $\OmegaRF/2\pi=78\pm20$\,kHz
 without having to determine $B_0$.

 he three experiments described below differ in the initial temperature, the atom number, the offset field $ (B_0) $, and the duration of the time-of-flight free expansion (1--10\,ms).

\subsubsection{A well-collimated pure atom laser:\label{sec:PureAtomLaser}} 

  \begin{figure}\center
 \includegraphics[width=0.55 \columnwidth]{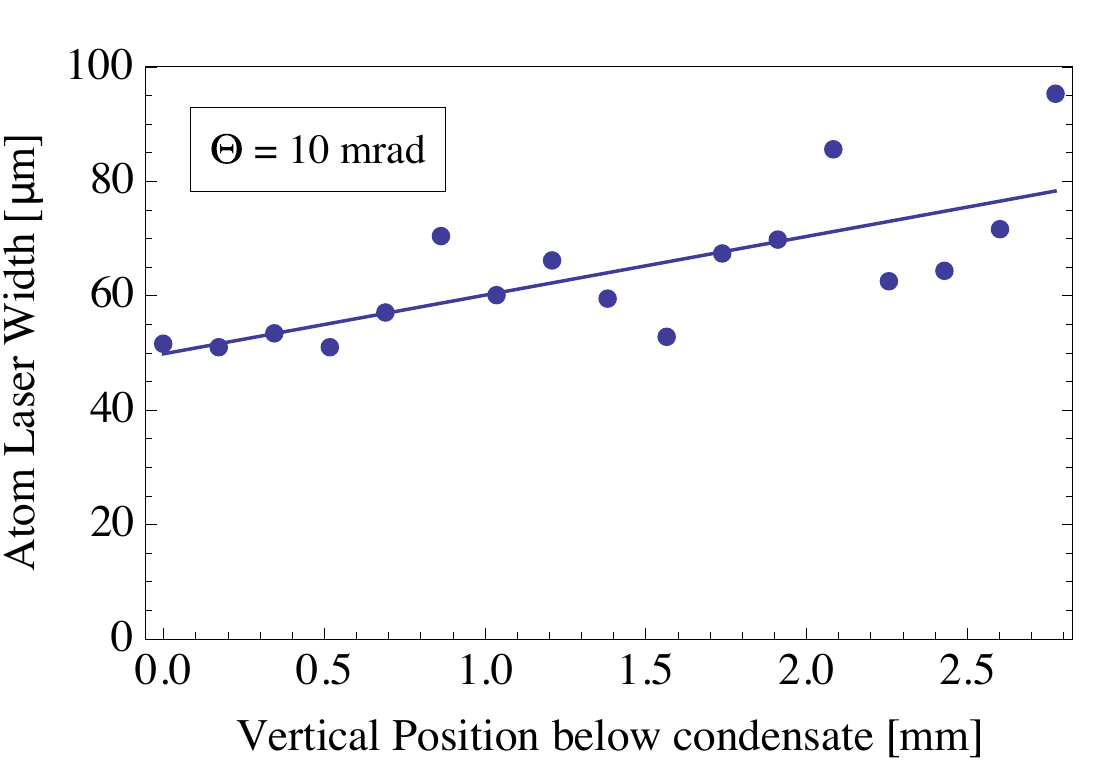} 
 \caption{
 The transverse size of the atom laser of the `pure' atom laser of Fig.\,\ref{fig:AtomLasers}a. The solid line is a straight line fit resulting in a divergence of 10\,mrad and an initial size of 50\,\textmu m. For comparison: A condensate containing all $3.1\times 10^4$ atoms detected in the atom-laser beam would have an axial Thomas--Fermi size of 44\,\textmu m in the absence of the dressing rf.\label{fig:pureLaserDivergenceReal}}
 \end{figure}

A well-collimated pure atom laser of $\sim 4.5$\,mm length and 2\,ms duration can be seen in Fig.\,\ref{fig:AtomLasers}a. 
It originated from a trap with non-dressed axial and radial trapping frequencies of 16.6\,Hz and 561\,Hz, respectively, and is therefore in the regime $\omegaZ\ll\Omega_{\rm oc}\ll\omegaR$. 
Since there are no observable thermal wings to the atom laser we fit only the atom-laser part of Eq.\,(\ref{eq:fitFunction}) and find a peak flux of $2.5\times 10^7$\,atoms per second for a total of $3.1\times 10^4$ atoms. 

As discussed in section \ref{sec:Divergence}, we expect the TDAP to be very well collimated. 
Fitting a straight line to the width as a function of distance (Fig.\,\ref{fig:pureLaserDivergenceReal}), we find a divergence of only $10$\,mrad, which compares well with the lowest divergences reported for any atom laser \cite{Jeppesen2008PRA, Le-Coq2001PRL}. 
The width of the atom laser at its origin is 50$\pm5$\,\textmu m.
A condensate containing all $3.1\times 10^4$ atoms detected in the atom-laser beam  would have an axial Thomas--Fermi radius of 44\,\textmu m. The agreement between the width of the atom laser beam and the Thomas--Fermi radius in the trap confirms that in the regime $\omegaZ\ll\Omega_{\rm oc}\ll\omegaR$ the condensate dynamics remain frozen in the axial direction and the outcoupling occurs over the full width of the condensate.

\subsubsection{A very high flux atom laser:}

 \begin{figure}\center 
 \includegraphics[width=0.55 \columnwidth]{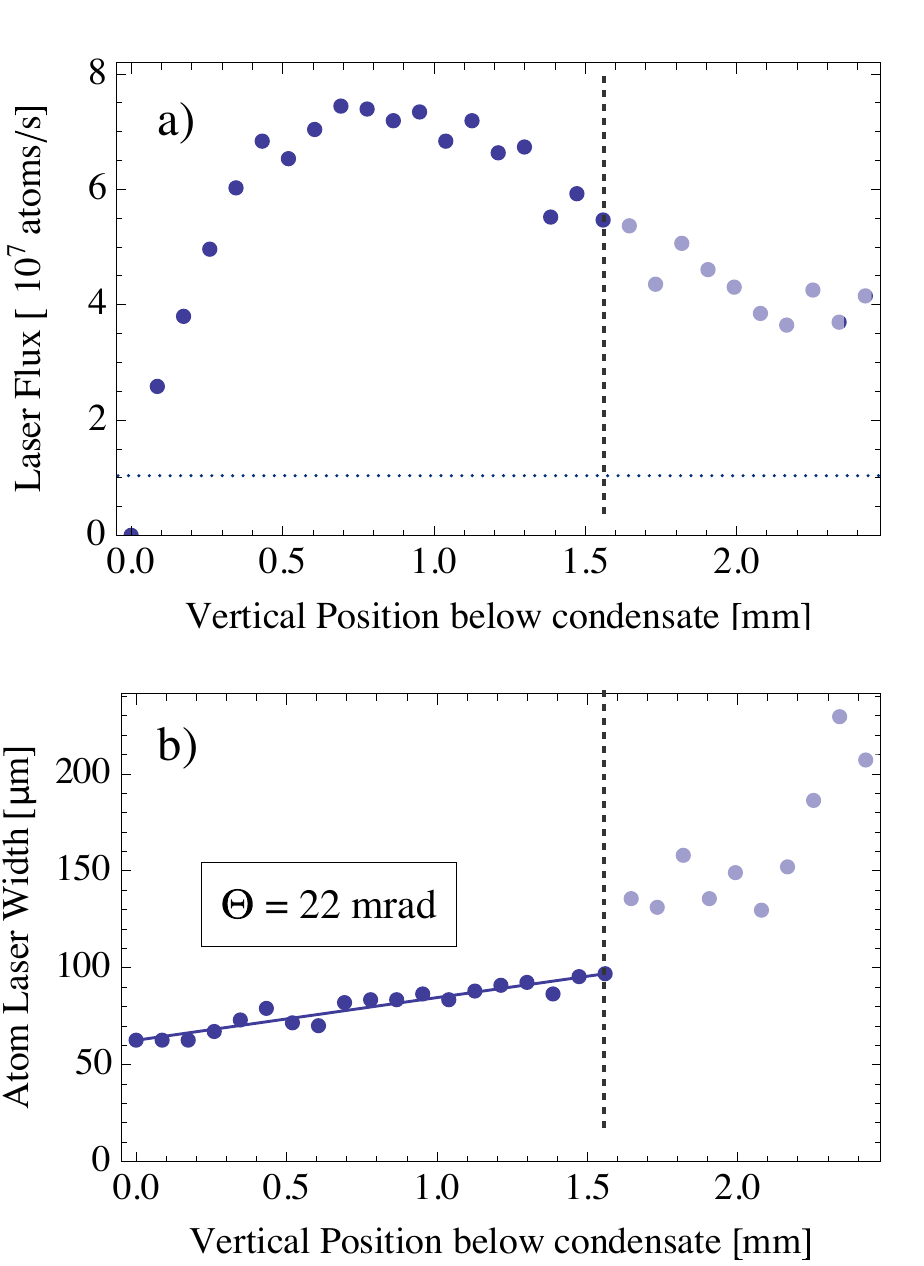}
 \caption{a) the atom flux  
  and  b) transverse size of the atom laser of Fig.\,\ref{fig:AtomLasers}b both plotted against the position below the condensate after time-of-flight expansion. 
  The flux and transverse sizes were determined from a fit of Eq.\,(\ref{eq:fitFunction}) (with $b=0$) to the vertically integrated slices of the absorption image (Fig.\ref{fig:AtomLasers}b). 
The horizontal dotted line shows the highest flux achieved previously \cite{Debs2010PRA}.
In b) at a position of 1.5\,mm below the condensate we observe  a step-change of the transverse size of the fit, which we attribute to the onset of the atom laser emission. The dashed vertical line serves as a guide to the eye. A linear fit to the atom-laser beam sizes (full line in b) results in a divergence of $\Theta = 22$\,mrad. The size at $y=0$ is 62\,\textmu m, which is very close to the Thomas--Fermi radius of 66\,\textmu m for $1.4\times 10^5$ atoms in the non-dressed trap (16.6\,Hz $\times$ 793\,Hz).\label{fig:highFlux} }
\end{figure}

Much effort has been focused on trying to maximise the flux available from a given BEC \cite{Robins2006PRL,Couvert2008EEL, Debs2010PRA}. 
The main obstacle to achieving very large fluxes from magnetically trapped BECs is the appearance at large coupling strengths $(\mbox{$\Omega^2_{\rm{rf}}$} \rightarrow \dot \Omega_{\rm L}/2\pi)$ of a bound state, which shuts off the atom laser \cite{Jeffers2000PRA,Robins2005PRA}. 
The reason for this is that  it is not possible by ramping up the rf-\emph{intensity} to adiabatically transfer the atomic population of a bare state  to a single dressed state.
In contrast to this, the TDAP atom laser relies on a strong rf, where the rf-\emph{frequency} is scanned. This  transfers the entire population adiabatically from the trapped to the anti-trapped state. We verified this in a Stern Gerlach experiment, where we found all atoms in the $m_{\rm F}=-2$ state after the rf-frequency had swept through the bottom of the trap. We also confirmed by rf-spectroscopy that the chemical potential of the original condensate matches the atom number detected in the atom laser beam.

\begin{figure}\center 
 \includegraphics[width=0.55\columnwidth]{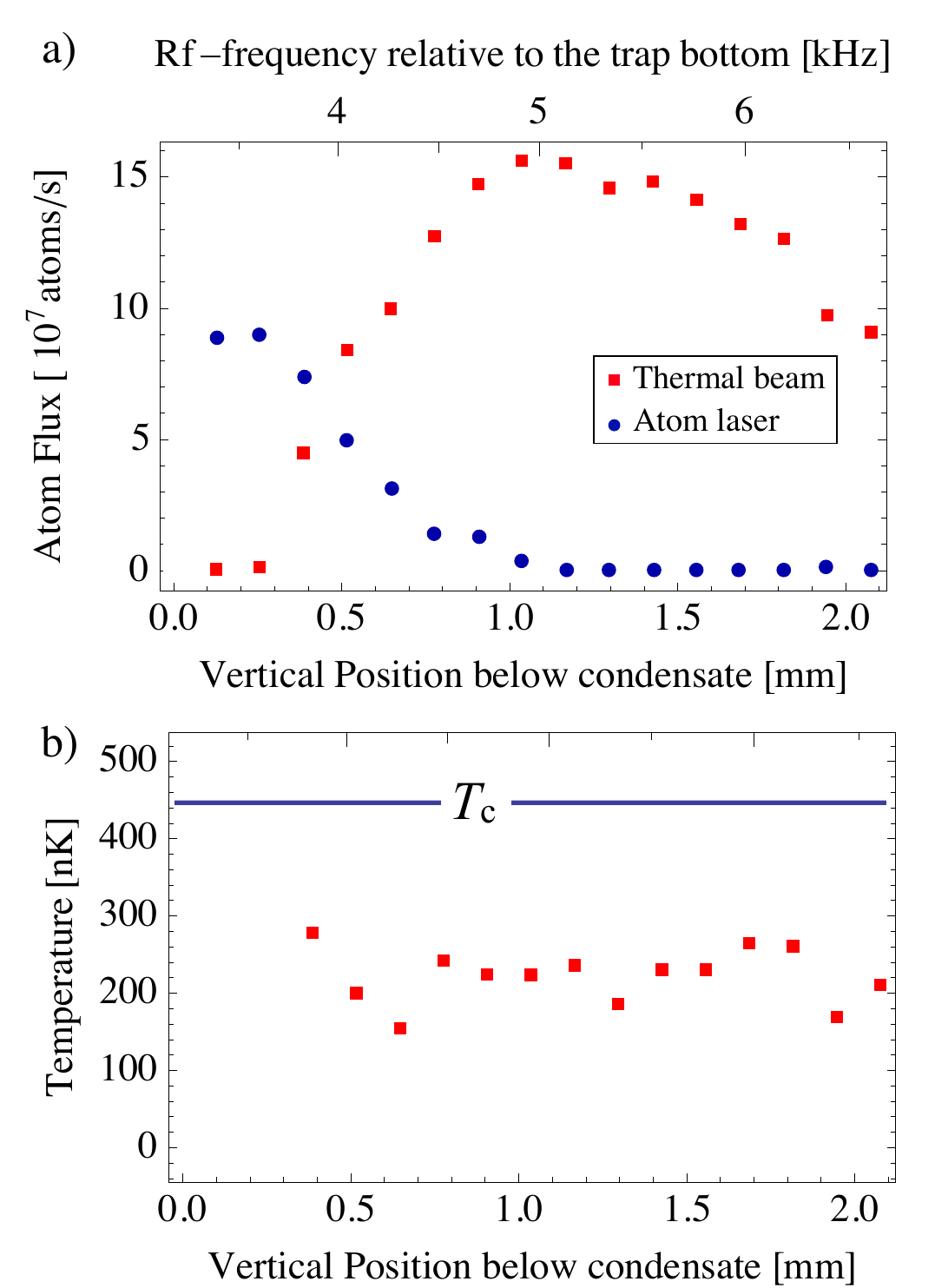} 
 \caption{a) Atom fluxes and b) temperatures of Fig.\,\ref{fig:AtomLasers}c.
 The horizontal axes are the position below the condensate after expansion (bottom) and the value of the rf-frequency at the time when the atoms were coupled out of the condensate (top). a)~shows the thermal and the coherent flux. At high rf-frequencies, only thermal atoms are outcoupled with the output coupling rate reaching a maximum of $1.6\times10^8$\,atoms/s. Atom laser emission sets in at about $5$\,kHz above the trap bottom and reaches a  maximum flux of $9\times 10^7$\,atoms/s. 
The thermal beam contains a total of $7.2\times 10^4$ atoms and the atom laser $2.1\times 10^4$ atoms. b)~shows the temperature of thermal component of the atom beam. 
 The solid blue line is the critical temperature for Bose-Einstein condensation in the trap in absence of the rf-dressing field for the total  number of atoms contained in the image ($9.3\times 10^4$ atoms). 
 \label{fig:CondTherm}} 
 \end{figure} 

The flux and transverse size of the atom laser of Fig.\,\ref{fig:AtomLasers}b can be seen in Fig.\,\ref{fig:highFlux}a) and b). 
The horizontal axis is the position below the condensate after a free expansion of 1\,ms. 
Atoms imaged at higher values of the position have been outcoupled at higher dressing frequencies $(\omegaRF)$. 
Since there are no discernible thermal wings, we fit the atom-laser part of Eq.\,(\ref{eq:fitFunction}) only. 
For smaller values of $\omegaRF$, i.e.\ smaller distances in Fig.\,\ref{fig:highFlux}b, we find a well collimated beam. 
At about 1.6\,mm, however, we find a sudden increase in the divergence of the beam.
The signal to noise ratio below this point did not permit us to analyse the exact shape of the atom beam.
The absence of thermal wings for lower rf-frequencies together with the sudden change in divergence of the atom beam indicate an onset of atom-lasing, i.e.~the instant when the trap depth becomes smaller than the chemical potential of the BEC. 

The flux in 
Fig.\,\ref{fig:highFlux}a reaches $7.4\times 10^7$\,atoms/s originating from only $9 \times 10^4$ atoms in the BEC. 
 This  is more than seven times larger than the previous maximum flux even with our initial BEC having only less than half as many atoms  \cite{Debs2010PRA}.  
 
\subsubsection{An ultra-cold thermal beam:\label{sec:thermalBeam}} 
Fig.\,\ref{fig:AtomLasers}c shows for the first time an atom beam, which contains concurrently an atom laser and a thermal atom beam. 
The upper part of Fig.\,\ref{fig:AtomLasers}c is an atom laser beam whereas the lower part is an ultra-cold thermal beam. 
In the central part of the figure the thermal and the atom laser overlap. 
We analyse the image by fitting the full Eq.\,(\ref{eq:fitFunction}) to the integrated image slices. 
A plot of the integrated slices and fits can be found in the supplementary material.  
We find $7.2\times 10^4$ atoms in thermal beam and $2.1\times 10^4$ atoms in the atom laser.
Fig.\,\ref{fig:CondTherm}a plots the thermal and atom laser fluxes against the position of the slice (bottom axis) and the value of the rf-frequency at the time of output coupling (top axis). 
As the rf-frequency ramps down (right to left on the plot), initially only thermal atoms are outcoupled with a peak rate of $1.5\times10^8$\,atoms/s. At an rf-frequency of about 5\,kHz above the trap bottom atom laser emission sets in. It reaches a flux of up to $9\times 10^7$\,atoms/s until the trap opens up completely. 
The divergence of the atom laser is 22\,mrad and its duration 0.35\,ms.

Fig.\,\ref{fig:CondTherm}b shows the temperature of the thermal beam as calculated from its width and expansion time. 
The solid line is the critical temperature of Bose-Einstein Condensation (450\,nK) in the non-dressed trap for the total number of atoms  both beams. 
The temperature of the thermal beam stays constant because at this point the collision rate is no longer sufficient to thermalise the atoms remaining in the trap. 
At only 200\,nK this is---to our best knowledge---the coldest thermal atom beam reported to date by more than two orders of magnitude \cite{Dieckmann1998PRA, Henson2012S, Grebenev1998S, Bartelt1996PRL}. Even though the absolute flux of the thermal atoms is very low compared to other atom beam sources, the extremely high energy resolution possible with this ultra-cold thermal beam will be useful e.g.\ in experiments comparing scattering of coherent and thermal atoms.

\vspace{1 cm}

\section{Conclusions}
We demonstrated a novel atom laser based on time-dependent adiabatic potentials. It allows an output coupling of the atoms at almost arbitrarily large rates and thus eliminates the bottleneck of the bound states in the traditional output couplers based on weak coupling. Our maximum atom laser flux of $7.4 \times 10^7$ atoms per second exceeds the maximum demonstrated anywhere by a factor of seven. Larger condensates and faster rf ramp rates will allow us to push this in the near future by another order of magnitude.

We observed for the first time an atom beam containing both a thermal and an atom-laser component. Its temperature of only 200\,nK is by more than two orders of magnitude the coldest thermal beam produced by any other technique. 

\section{Acknowledgments} 
We acknowledge the financial support of the Future and Emerging Technologies (FET) programme within the 7th Framework Programme for Research of the European Commission, under FET grant number: FP7-ICT-601180 and of the Marie Curie Excellence s under Contract MEXT-CT-2005-024854.
V.B. gratefully acknowledges that her research has been co-financed by the European Union (European Social Fund ESF) and Greek national funds through the Research Funding Program: Heracleitus II\@. N.K.E. is partially supported by the Research Project ANEMOS co-financed by the European Union (European Social Fund - ESF) and Greek national funds through the Operational Program ``Education and Lifelong Learning'' of the National Strategic Reference Framework (NSRF) - Research Funding Program: Thales. 
Finally, we would like to thank J.\,Close and T.P.\,Rakitzis for encouragement and critical comments.
\clearpage
\providecommand{\newblock}{}

\cleardoublepage
\section{APPENDIX}
%
 
\subsection{Flux Calculation \label{sec:VelocityMapping}}
Near the trap bottom $ ( y \ll B_0) $ we can write the trapping potential in the direction of gravity $ (x=0, z=0) $ as
\begin{equation}
V (y) =m_{\mbox{\tiny{F}} } \hbar \sqrt{\left (\frac{\alpha^2 _{\omega } y^2}{2 \OmegaL}-\Delta \omegaRF\right) ^2+\Omega_{\rm rf}^2}\,+M g_{\rm e\,} y\,, \label{eq:potenitalHarm}
\end{equation}
where $\Delta \omegaRF= \omegaRF -\left|\gF \right|\muB B_0 /\hbar $ is the detuning of the rf-frequency from the trap bottom in the absence of the dressing field, and $\alpha _{\omega }$ is the gradient of the radial qua\-dru\-pole field written in angular frequency units
 $ (\alpha _{\omega }=\gF \muB \alpha /\hbar) $.
If the rf-field is linearly polarised and orthogonal to the z-axis, then close to the centre of an elongated Ioffe--Pritchard trap the coupling strength can be simplified to $\OmegaRF = \left|\gF \right|\muB B_{\rm rf}/ (2\hbar) $. 
We calculate the trap depth $ (\Delta V_{\rm trap}) $ by taking the difference $(\Delta V_{\rm trap} = V_{\rm{out}} - V_0)$ of the trap minimum $(V_0)$ and output coupling point $(V_{\rm{out}})$. 

\begin{figure}[h!]
 \center
\centering
\includegraphics[width=0.55\columnwidth]{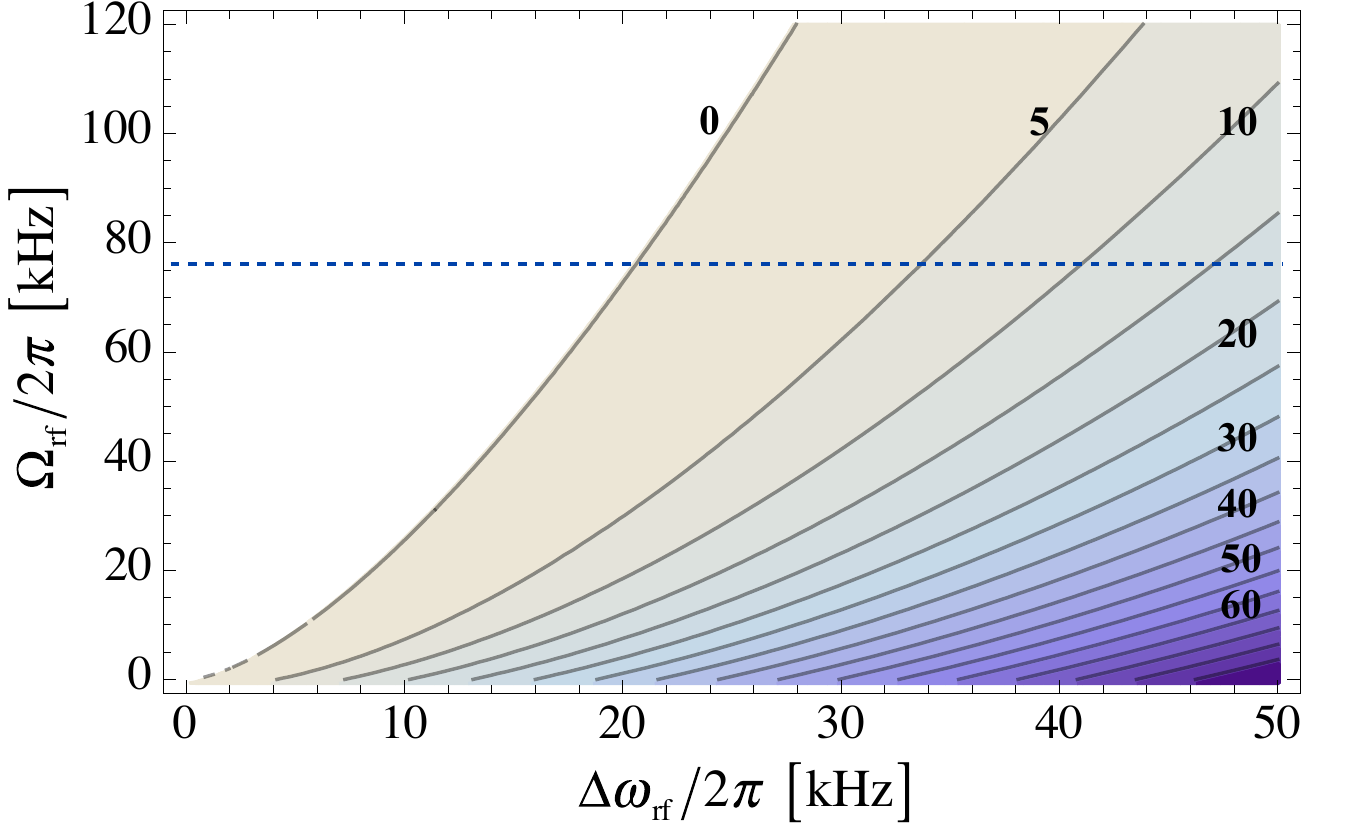}
\caption{The trap depth $ \Delta V_{\rm trap}$ as a function of 
 $\Delta \omegaRF$ and $\OmegaRF$. In the white region there is no stable trap. The numbers inside the plot denote the values of the corresponding contour line of $\Delta V_{\rm trap}$ in kHz. The horizontal dashed line correspond to $\OmegaRF/2\pi=76$\,kHz, which is the value used in the experiments presented here. 
}
\label{fig:trapDepth2D} 
\end{figure}

Fig.\,\ref{fig:trapDepth2D} shows the trap depth for our typical experimental parameters as a function of the coupling strength $\OmegaRF$ and of the rf-frequency relative to the trap bottom. Note that because of gravity and the weakening of the trap by the dressing field, the depth of shallow traps is much smaller than the detuning of the rf-frequency from $B_0$ $(\Delta V_{\rm trap}\ll \hbar\, \Delta \omegaRF)$.

During the lasing phase the atoms are accelerated both by gravity $(g_{\rm e})$ and by the gradient of the magnetic field $(a_\alpha=\mF\gF \muB \alpha /M)$. 
During the time-of-flight expansion they feel only earth's acceleration. 
The atoms in the slice located at position $x$ of the image have been accelerated by the magnetic gradient for a time:
\begin{equation}
t_{\mbox{\tiny L}}= \sqrt { {t_{\mbox{\tiny E}}}^2+\frac{{2x}}{{{g_e} + {a_\alpha }}}} - {t_{\mbox{\tiny E}}}\,. \label{eq:LasingTime}
\end{equation}
For convenience, we set $x$=0 at the position of the BEC after time-of-flight expansion of duration $(t_{\mbox{\tiny E}})$ .
The velocity of the atoms just after the time-of-flight expansion is given by 
$
v_{\mbox{\tiny E}}= (g_{\rm e}+a_{\alpha})\,t_{\mbox{\tiny L}} + g_{\rm e}\,t_{\mbox{\tiny E}} 
$. The flux just after the lasing phase but before the time-of-flight expansion is then
\begin{equation} j(x)=n_{\mbox{\tiny 1D}} \left(a_\alpha + g_e\right)\left(t_{\mbox{\tiny E}}+t_{\mbox{\tiny L}}\right)\,,\label{eq:FluxFromPosition}
 \end{equation}
 where $n_{\mbox{\tiny 1D}}$ is the one dimensional density of the atoms, i.e.\ the number of atoms conatained in a slice divided by its thickenss.
\subsection{Temperatures \label{sec:Temperature}}
The temperature of a slice of atoms is calculated from the fit of Eq.\,(\ref{eq:fitFunction}) as 
\begin{equation}
T=\frac{1}{2 k_{\mbox{\tiny{B}}}}M\!\left(\frac{\Delta x_{\rm t}} {t_{\mbox{\tiny{L}}}+t_{\mbox{\tiny{E}}}}\right)^2\label{eq:temperature},
\end{equation}
where $ k_{\mbox{\tiny{B}}}$ is the Boltzman constant. 
This slightly overestimates the temperature, because it does not take into account that  the atoms are `anti-trapped', i.e.~it neglects the contribution to the $\Delta x_{\rm t}$ that originates during the lasing phase from the negative curvature of the potential of Eq.\,(\ref{eq:potenital}).

\subsection{Beam Divergence \label{sec:Divergence}}
  \begin{figure}\center
 \includegraphics[width=0.55 \columnwidth]{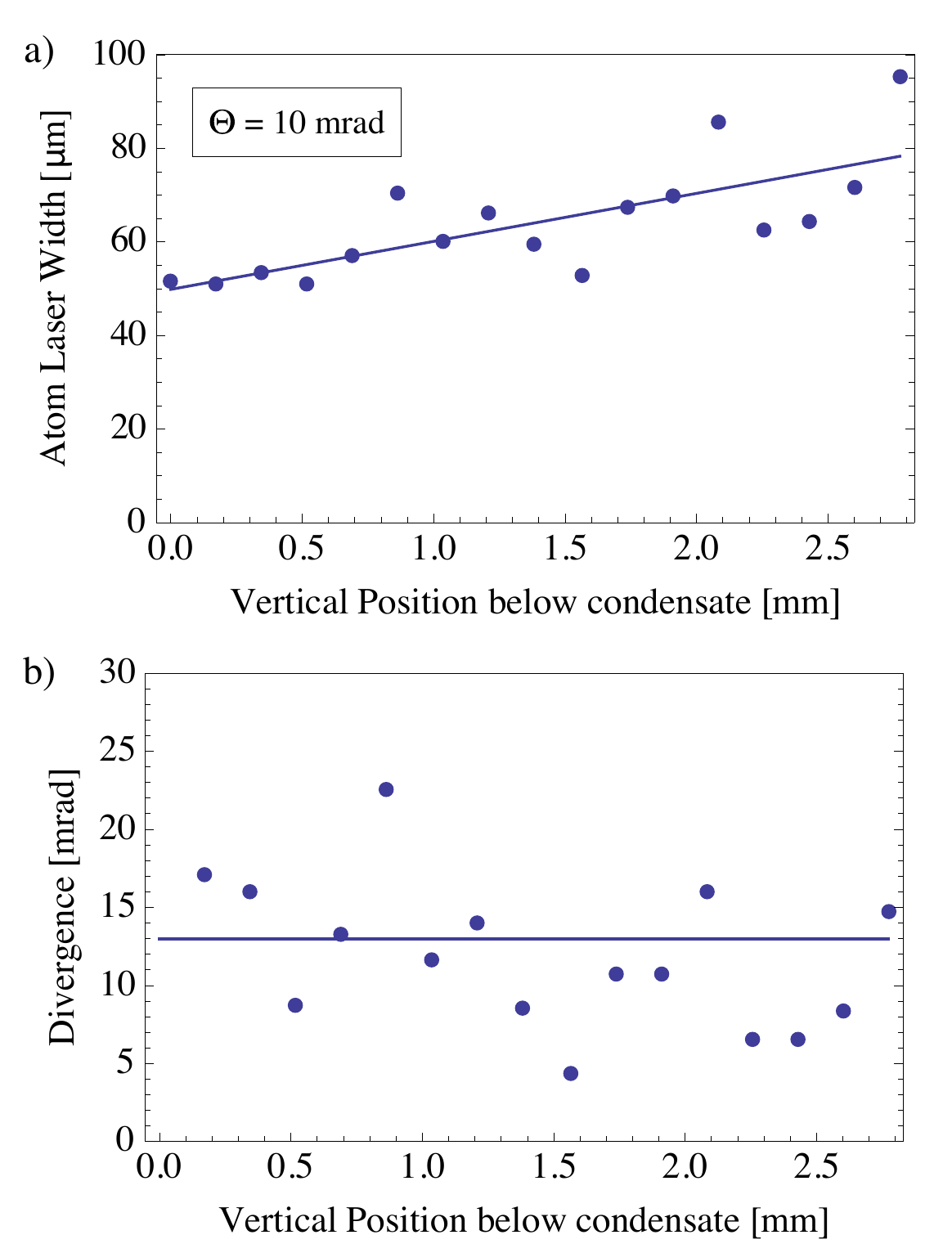} 
 \caption{Laser width and divergence of the `pure' atom laser of Fig.\,\ref{fig:AtomLasers}a in real and momentum space. a) the transverse size of the atom laser with the solid line being a straight line fit (for convenience repeated from Fig.\,\ref{fig:pureLaserDivergenceReal}). b) the divergence of the atom laser based on the velocities assuming as an initial size the Thomas--Fermi radius of 37000 atoms in a trap of $16.5\times 561$\,Hz trapping frequency (44\,\textmu m). The solid line is the mean divergence (13\,mrad).
 \label{fig:pureLaserDivergence}}
 \end{figure}
 
One can determine the divergence $\Theta$ of an atom laser beam either in real space or in momentum space. 
In real space one take the arctangent of the slope of a straight line fit to the size of the atom laser beam $(\Delta x_{\rm c})$ versus the position x  (Fig.\,\ref{fig:pureLaserDivergence}a). 
In momentum space we take the arctangent of the calculated velocities for each slice separtely (Fig.\,\ref{fig:pureLaserDivergence}b). The latter method requires knowledge of the initial size of the beams, which we estimate as the Thomas--Fermi radius of a BEC in the non-dressed trap containing all the atoms detected in the atom laser. The velocities are calculated from the position in the image on the basis of the gravitational and magnetic acceleration.

The results are not limited by the imaging resolution is 5\,\textmu m nor by the pixel size of 43\,\textmu m. The atoms in image slices from different positions have been outcoupled at different lasing times and might therefore have had different initial sizes. A more thorough analysis of the divergence would therefore have to follow the expansion of a single image slice.

For the `pure' laser (Fig.\,\ref{fig:pureLaserDivergence}) the divergence in found a divergence of $10\pm3$\,mrad in real and $13\pm6$\,mrad in momentum space. 
For the `large flux' laser we found a divergence of $22\pm1$\,mrad in real space (Fig.\,\ref{fig:highFlux}b) and $13\pm2$\,mrad in momentum space (for the region of sufficient signal to noise, 0.7--13\,mm). 

\subsection{Numerics \label{sec:numerics}}

We consider the dynamics of the Gross-Pitaevskii equation
\begin{equation}
i\hbar\frac{\partial\psi}{\partial t}=
-\frac{\hbar^2}{2m}\nabla^2\psi
+V_d(y,z)\psi+
\frac{4\pi\hbar^2}{m} a_s|\psi_1|^2\psi
\end{equation}
with a dressed potential 
\begin{equation}
 V_d=\sqrt{
 \left( \hbar \OmegaRF \right)^2
 +\left[
 \mu
 \sqrt{( \alpha y)^2+\left(B_0+\frac{\beta }{2}z^2 \right)^2} -\hbar \omegaRF \right]^2} \,\,+g_{\rm e}\, M\, y.
\end{equation}
In order to reduce the computational complexity of the problem, we consider only the two dimensional problem. 
We introduce the dimensionless quantities $\tau=t/t_0$, $\xi=y/y_0$, $\zeta=z/y_0$, and $\psi=A\phi$ and then set the kinetic term coefficient to $1/2$ and the nonlinear coefficient to unity by requiring that
$y_0=\sqrt{\hbar\,t_0/M}
$ 
and
$
A=
\sqrt{M/(4\pi \hbar t_0 a_s)}$.
The resulting normalised Gross-Pitaevskii equation has the form
\begin{equation}
i\frac{\partial\phi}{\partial\tau}=
-\frac{1}{2}\nabla^2\phi
+V(\xi,\zeta)\psi+
\gamma|\psi_1|^2\psi
\label{eq:nls2}
\end{equation}
with a potential
\begin{equation}
V=\sqrt{
\kappa^2
+\left[
\sqrt{(\alpha_y y)^2+\left[F+ (\alpha_zz)^2 \right]^2} - \Delta(\tau) 
\right]^2}+Gy
\label{eq:pot2}
\end{equation}
and we have defined $\kappa=\OmegaRF t_0$, 
$\alpha_y=t_0\muB\alpha y_0/\hbar$,
$F=B_0t_0\muB/\hbar$,
$\alpha_z=t_0\muB\beta y_0^2/(2\hbar)$,
$G=gMy_0t_0/\hbar$,
$\Delta(\tau) = 2\pi \omegaRF(t)(\tau t_0)t_0$.

Using the numerical values 
$M=1.44\times10^{-25}\unit{kg}$, 
$g_{\rm e}=9.81\unit{m/s^2}$, 
$m_{\mathrm{F}}=2$, 
$g_\mathrm{F}=1/2$, 
$\mu=g_\mathrm{F}m_\mathrm{F}\muB
=0.73\unit{MHz}$,
and selecting
$\alpha=8 \unit{T/m}$, $\beta=3\times10^5\unit{T/m^2}$, $B_0=0.5\times10^{-4}\unit{T}$, 
$\OmegaRF=2\pi\,20\unit{kHz}$, 
and
$t_0=1/(2\pi\,5000)\unit{s}$, we derive the parameters for the normalised potential. In particular, 
the space and density scalings are given by
$y_0=1.5\times10^{-7}\unit{m}$, 
$A=2.44\times10^{10}\unit{m^{-3/2}}$,
while the potential parameters are 
$G=0.065 
$, $\alpha_y=3.41
$, $F = 140
$, $\alpha_z =0.099 
$, $\kappa = 4$. To release the condensate from the trapping potential we set the initial value of the frequency to $\omegaRF=4.6\times10^6\unit{Hz}$ and gradually decrease its value. The initial state of the condensate is given by the Thomas--Fermi approximation. In the initial states of the dynamics (up to $\tau=120$) we have $\Delta'(\tau)=-1.75
$ while for $\tau>120$ this value is increased to $\Delta'(\tau)=-0.035$. Thus for $\tau>120$, after a ``relaxation'' of the BEC dynamics of the condensate, it becomes adiabatic and quasi-stationary in the observed time scale. 

 \begin{figure}\center
\centerline{\includegraphics[width=0.8\columnwidth]{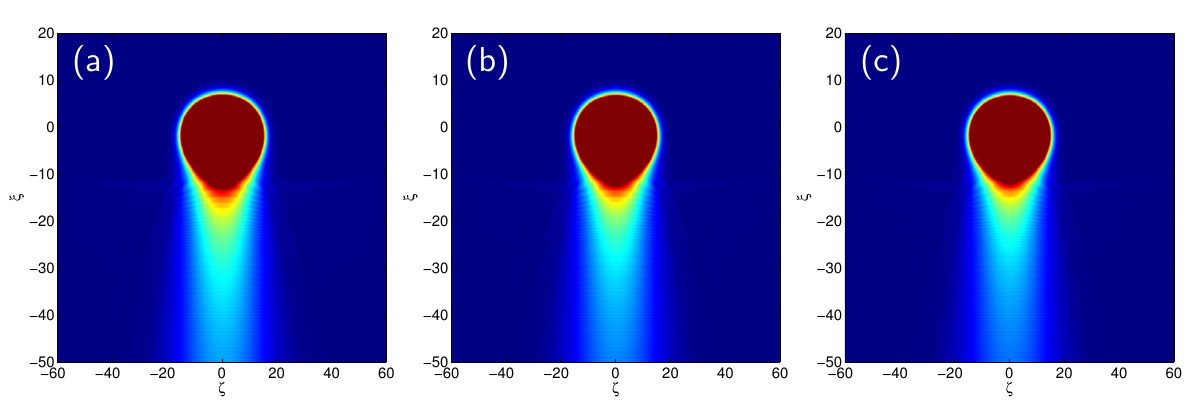}}
\caption{Density plots of the atom laser  for evolution times $\tau=500,600,700$.\label{fig:AtomLaserNumerics}}
\end{figure}

In Fig.~\ref{fig:AtomLaserNumerics}, we see the dynamics of an atom laser BEC for different evolution times. 
We used a forth order split step Fourier method to computationally solve Eqs.\,(\ref{eq:nls2})-(\ref{eq:pot2}). The scheme involves periodic boundary conditions, and thus, in order to ``absorb'' the wave close to the boundaries we introduced a boundary layer with linear loss. 
In Fig.~\ref{fig:AtomLaserNumerics} we note that there are only very small differences in the density profiles for $\tau=500, 600, 700$ and thus the dynamics are adiabatic and quasi-stationary. The small changes in the dynamics observed in the figure are attributed to small changes of the potential with time. 

\clearpage

\section{supplementary material}
\begin{enumerate}
\item A grayscale version of Fig.\,\ref{fig:AtomLasers}.
\item The slices and fits used in the analysis of Fig.\,\ref{fig:AtomLasers}c, i.e.\,the atom beam containing both a thermal beam and an atom laser beam.
\item A video of the numeric simulation of an atom laser.
\end{enumerate}

\cleardoublepage

\end{document}